\documentstyle[pra,aps,twocolumn,floats,psfig]{revtex}

\begin{document}

\draft

\title{Collisions of cold magnesium atoms in a weak laser field}

\author{Mette Machholm~\cite{Mette}$^{1,2}$, Paul S. Julienne$^3$, and
Kalle-Antti Suominen$^{2,4}$}

\address{$^1$Department of Chemistry, Aarhus University, Langelandsgade 140,
DK-8000 Aarhus C, Denmark\\
$^2${\O}rsted Laboratory, NBIfAFG, University of Copenhagen,
Universitetsparken 5, DK-2100 Copenhagen {\O},
Denmark\\
$^3$National Institute for Standards and
Technology, 100 Bureau Drive, Stop 8423, Gaithersburg, MD 20899-8423\\
$^4$Helsinki Institute of Physics, PL 9, FIN-00014 Helsingin yliopisto,
Finland}

\date{\today}

\maketitle

\begin{abstract}
We use quantum scattering methods to calculate the light-induced collisional
loss of laser-cooled and trapped magnesium atoms for detunings up to 30
atomic linewidths to the red of the $^1$S$_0$-$^1$P$_1$ cooling
transition.  Magnesium has no hyperfine structure to complicate the theoretical
studies.  We evaluate both the radiative and nonradiative mechanisms of trap
loss.  The radiative escape mechanism via allowed $^{1}\Sigma_{u}$ excitation
is dominant for more than about one atomic linewidth detuning.   Molecular
vibrational structure due to photoassociative transitions to bound states
begins to appear beyond about ten linewidths detuning. 

\end{abstract}

 
\bigskip

\narrowtext

Light-induced collisions between cold, neutral alkali atoms have been 
widely studied experimentally and theoretically in magneto-optical 
traps~\cite{exp2,theory2,RMP}.  Such collisions cause loss of atoms 
from the trap due to radiative or nonradiative molecular processes 
after laser excitation of molecular states of the atom pair at large 
internuclear separation $R$.  These loss processes for alkali atoms 
are still not well-understood when the laser detuning $\Delta$ is only 
a few natural atomic linewidths $\Gamma_{\rm at}$ to the red of atomic 
resonance, primarily because of complications due to the molecular 
hyperfine structure in the alkali species~\cite{RMP,Fioretti97}.  Molecular 
vibrational and rotational structure in the spectrum of loss versus 
$\Delta$ is unresolved due to broadening by fast spontaneous decay and 
the high density of molecular states.  On the other hand, high 
resolution photoassociation spectra for trap loss at much larger 
detunings are quite well-understood quantitatively, even for alkali 
species~\cite{RMP,Lett95}.

The study of trap loss collisions in alkaline earth species offers a 
significant opportunity to improve our understanding of 
collisional loss processes in the small detuning regime.  
The major isotopes of Be, Mg, Ca, Sr, and Ba have no nuclear spin, and 
thus lack hyperfine structure.  Alkaline earth atoms can also be 
cooled and trapped (e.g.~Mg~\cite{Ertmer93}, Ca~\cite{Kurosu90}, 
Sr~\cite{Kurosu90,Katori99}).  They offer an excellent and previously 
rather unexplored testing ground for cold collision theories.  A 
recent study of Sr has taken a first step in this 
direction~\cite{Dinneen99}.  One intriguing aspect of the alkaline 
earth collisions is the possibility to observe the effect of 
relativistic retardation corrections on the trap loss dynamics by 
exciting a molecular state with a dipole-forbidden transition at small 
R which becomes allowed at large R. This predicted effect for the $2_u$ 
state of alkali species~\cite{JV91} is masked by other states, but may 
be prominent for the $^1\Pi_g$ state of alkaline earth species.

\begin{figure}[htb]
\noindent\centerline{
\psfig{width=80mm,file=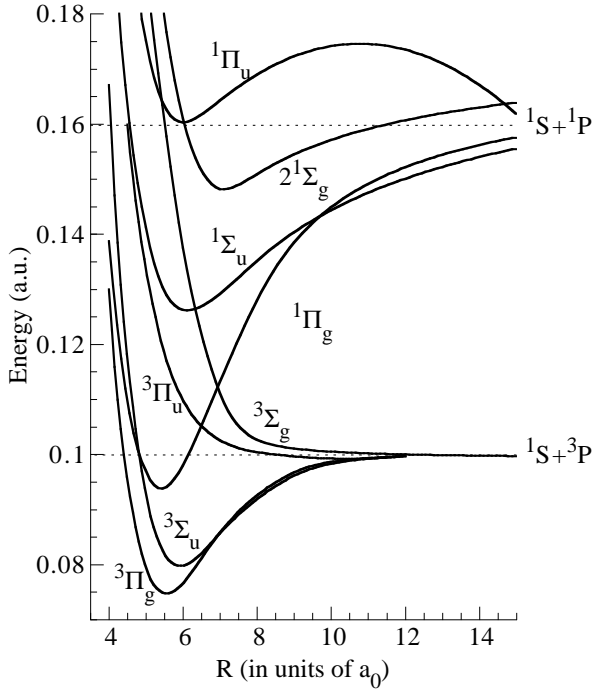}
}
\caption[f1]{The molecular states of Mg$_2$ in atomic units (1 a.u. = 
4.3597482$\times 10^{-18}$ J) corresponding to the asymptotic atomic 
states $^1$S$_0$+$^1$P$_1$ and 
$^1$S$_0$+$^3$P$_{0,1,2}$~\cite{Stevens77}; the zero of energy is at 
the ground state $^1$S$_0$+$^1$S$_0$ asymptote.  There 
are four states related to each asymptote, of which two are attractive 
and two are repulsive at large $R$, where the system is expected to be 
resonant with the laser field.  Each set of two states has one $\Pi$ 
state and one $\Sigma$ state.
\label{potgraph}}
\end{figure}

In this paper we calculate collisional loss rate coefficients of Mg 
atoms in a magneto-optical trap, induced by a weak laser field detuned 
up to 30$\Gamma_{\rm at}$ to the red of atomic resonance ($\Gamma_{\rm 
at}=2\pi\cdot 80$ MHz).  This spans the range from small detuning, 
where the spectrum is unstructured, to the far-detuning range, where 
resolved photoassociative features due to specific molecular 
vibrations begin to appear.  For $^{24}$Mg we assume that the cooling 
transition is $^1$S$_0$-$^1$P$_1$, ($\lambda=285.2$ nm).  
Fig.~\ref{potgraph} shows the short-range part of the Mg$_2$ 
potentials~\cite{Stevens77}.  A laser with $\Delta\sim$ 1-30 
$\Gamma_{\rm at}$ excites molecular states with attractive potentials 
at some Condon distance $R_C\gtrsim 250\ a_{0}$ ($a_{0}$ = 0.0529177 
nm).  In $^{24}$Mg these states can be either the $^1\Sigma_u^+$ state 
or the $^1\Pi_g$ state.  Spin-orbit coupling $H_{\rm SO}$ mixes 
$^1\Pi_g$ with the $^3\Sigma_g$ state at a curve crossing around 
$R\sim 7$~$a_{0}$ and mixes $^1\Sigma_u^+$ with $^3\Pi_u$ around 
$R\sim 5$~$a_{0}$.  After a change of molecular state the 
quasimolecular atom pair can separate to a $^1$S$_0$ atom and a $^3$P 
atom ($^3$P$_0$, $^3$P$_1$ or $^3$P$_2$), with a shared kinetic energy 
increase equal to about 0.06 atomic units ($\Delta E/k_B\sim$ 19 000 
K).  These hot atoms can not be recaptured by the trapping and cooling 
lasers and they are thus lost from the trap.  In alkalis this process 
arises from collisional transfer between two fine structure states of 
the same electronic configuration, and is called the fine structure 
changing mechanism (FS).  Here we call it the state changing 
mechanism (SC) between different electronic states.

%fig1 here

After excitation at $R_C$ the quasimolecule obtains kinetic energy via 
internal acceleration by the attractive potentials.  If it decays back 
to the ground state, it preserves this gain in kinetic energy, thereby 
increasing the kinetic energy of the asymptotic collision products 
(two $^1$S$_0$ atoms).  If the energy gain exceeds the trap depth, the 
atoms can escape from the trap.  This is the radiative escape 
mechanism (RE).  Spontaneous emission has a crucial role in both the 
SC and RE loss mechanisms, and has to be included in the theoretical 
treatment.

The structure of the quasimolecule at large $R$ is determined by the 
dipole-dipole interaction, for which we use the long-range retarded 
potentials from Ref.~\cite{Meath68}.  These join with the short-range 
wells in Fig. (\ref{potgraph}) to give potentials that support a 
number of bound vibrational states of the quasimolecule.  The 
molecular linewidths with relativistic retardation corrections for the 
two relevant excited states with attractive potentials 
are~\cite{Meath68}
\begin{eqnarray}
   \Gamma_{\Sigma_u}(R) &=& \Gamma_{\rm at} \left\{1-\frac{3}{u^3}\left[
                       u\cos(u)-\sin(u)\right]\right\}, \\ \label{Gs}
   \Gamma_{\Pi_g}(R)    &=& \Gamma_{\rm at} \left\{1-\frac{3}{2u^3}\left[
                       u\cos(u)-(1-u^2)\sin(u)\right]\right\} ,
\end{eqnarray}
where $u=R/\lambdabar$, and $\lambdabar=\lambda/2\pi=857.7\ a_{0}$.  
For small $u$ we get $\Gamma_{\Sigma_u} \simeq \Gamma_{\rm at} 
(2-u^2/10)$ and $\Gamma_{\Pi_g} \simeq \Gamma_{\rm at} u^2/5$.  Note 
that the optical coupling between the quasimolecule and the laser 
field is proportional to $\sqrt{\Gamma_{A}}$ ($A=\Sigma_u,\Pi_g$).  As 
detuning $\Delta$ increases the probability to excite the $^1\Pi_g$ 
state is reduced.  On the other hand, the survival of the 
quasimolecule on the $^1\Pi_g$ state is much better than for the 
$^1\Sigma_u$ state.  We expect that RE loss comes mainly from the 
$^1\Sigma_u$ state, and that qualitatively the $^1\Pi_g$ state and the 
$^1\Sigma_u$ state contributions to SC loss behave differently as a 
function of $\Delta$.  The important question is the relative 
magnitudes of the various loss mechanisms.

The standard small-detuning studies of trap loss ignore the 
vibrational structure of the trap loss, and consider the problem as a 
fully dynamical one, involving only the electronic 
potentials~\cite{theory2}.  This is based on the expectation that at 
small detunings the vibrational structure can not be resolved.  In the 
time-dependent picture this simply means that after the excitation at 
large $R$ the quasimolecule decays well before it completes one or 
more vibrations.  However, here the retardation effect opens the 
$^1\Pi_g$ state for excitation but closes it quickly for decay, so we 
might expect to see vibrational structure in trap loss even at small 
detunings if the $^1\Pi_g$ state SC contribution dominates over the 
$^1\Sigma_u$ state SC and RE contributions.

The ground state potential of Mg$_{2}$ is essentially flat in the 
long-range region of excitation.  Since a number of total angular 
momentum $J$ values contribute to trap loss in our detuning range, we 
assume that effects of the unknown phase shifts in the ground state 
wavefunctions due to the short-range potential is removed by summing 
over $J$ and thermal averaging.  Thus, we represent the ground state 
potential by a Lennard-Jones 6-12 potential with a well depth of 
0.0028 atomic units and an inner turning point of 
6.23~$a_{0}$~\cite{Stevens77}.  Since the excited {\it ab initio} 
potentials do not permit a calculation of vibrational levels to 
spectroscopic accuracy, the specific forms of the short-range excited 
state potentials are not important for our purposes of modeling the 
qualitative structure and magnitude of the collisional loss.  We fit 
Lennard-Jones 3-6 potentials to the {\it ab initio} potentials of 
Ref.~\cite{Stevens77}, keeping the long-range form fixed to its known 
value.  Modeling of experimental data should ultimately permit 
calibration of the unknown phases associated with the uncertain inner 
parts of the potentials.

A complete study must include the three-dimensional aspects of 
molecular rotation.  Since all collision directions are possible 
relative to laser polarization, the effect of axis rotation is to 
introduce the usual molecular branch structure of the possible 
transitions.  The angular momentum $J''$ in the ground state can only 
be that of axis rotation, or partial wave $\ell'' = J''$.  In the 
excited state $^1\Sigma$, $^1\Pi$ molecular basis, $\ell$ is not a 
good quantum number but total angular momentum $J'$ is.  The possible 
transition branches are P, Q, and R, for which $J'$ = $J''-1$, $J''$, 
and $J''+1$ respectively.  The quasimolecule ground state can couple 
to the $^1\Sigma_u$ state only by P and R branches, but to the 
$^1\Pi_g$ state by P, Q and R branches.  The radiative coupling terms 
for the two electronic states for each branch and partial wave 
$\ell''$ are shown in Table~\ref{tbl1}.

%table here
\begin{table}[htb]
\begin{tabular}{dddd}
State $A$ & Branch $B$ & $l''=0$ & $l''\neq 0$\\
\hline
$\Sigma$ & P & 0            & $\sqrt{l''/3}$      \\
$\Sigma$ & R & $\sqrt{2/3}$ & $\sqrt{(l''+1)/3}$  \\
$\Pi$    & P & 0            & $\sqrt{(l''-1)/3}$  \\
$\Pi$    & Q & 0            & $\sqrt{(2l''+1)/3}$ \\
$\Pi$    & R & $2/\sqrt{3}$ & $\sqrt{(l''+2)/3}$  \\
\end{tabular}
\vskip0.5cm
\caption[t1]{The radiative coupling weight factors $\alpha_{A,B,l''}$ for each
electronic state, ground state partial wave $l''$ and excitation branch,
obtained by summing over the degenerate projection states. The full radiative
coupling in atomic units is $V_{A,B,l''}=1.16\times 10^{-4}\alpha_{A,B,l''}
\sqrt{I\ (\mbox{W/cm}^2)} \sqrt{\Gamma_A(R)\ (\mbox{a.u.})}$.
\label{tbl1}}
\end{table}

Since the laser field is assumed to be weak, reexcitation of any 
decayed quasimolecular population can be ignored and loss rates can be 
calculated by the complex potential 
method~\cite{Boesten93,Julienne94}.  Additionally, each ground state 
partial wave only couples to the rotational states of the $^1\Sigma_u$ or 
$^1\Pi_g$ state, but these states do not couple further to other ground 
state partial waves.  Thus, the weak field approximation permits us to 
truncate the problem into a set of independent three-state problems 
for each initial partial wave with one ground state $g$, one excited 
state $e$, and one probe state $p$.  The total thermally averaged loss 
rate coefficient via state $e$ is
\begin{eqnarray}
   K(\Delta,T) &=& \frac{k_BT}{hQ_T} \int_{0}^{\infty}
   \frac{d\varepsilon}{k_BT}\exp\left(
   -\frac{\textstyle\varepsilon}{\textstyle k_BT}\right)
   \sum_{l''_{\rm even}}   (2l''+1)
  \nonumber\\
   &&\times\sum_{B=P,Q,R} |S_{gp}(\varepsilon,\Delta,l'',B)|^2.
   \label{K2}
\end{eqnarray}
where $\varepsilon$ is the relative asymptotic initial kinetic energy, 
$Q_T=(2\pi\mu k_BT/h^2)^{3/2}$ is the translational partition function 
($\mu$ is the reduced mass of the two colliding atoms), and 
$|S_{gp}|^{2}$ represents the probability of a transition to a probe 
channel $p$ which simulates the effect of the SC or RE exit channels 
\cite{Julienne94}.  Identical particle exchange symmetry ensures that 
only even partial waves exist for the ground state.  The ground state 
centrifugal potential provides a natural cutoff $\ell_{max}$ to the 
sum over $l''_{\rm even}$ in Eq.~(\ref{K2}) where 
$\hbar^2\ell_{max}(\ell_{max}+1)/2\mu R_t^2>\epsilon$ at the classical 
turning point $R_t$ for collision energy $\epsilon$.

We calculate $S_{gp}(\varepsilon,\Delta,l,B)$ using a complex
potential~\cite{Boesten93,Julienne94}, i.e.,  spontaneous decay is
taken into account by adding an imaginary decay term
$-i\hbar\Gamma_A(R)/2$ ($A=\Sigma_u,\Pi_g$) to the excited state
potentials.  We solve the corresponding time-independent
Schr{\"o}dinger equation with the appropriate boundary conditions and
obtain the $S$-matrix elements.

For the SC loss studies the probe state represents the actual loss 
channel due to the curve crossings mentioned above which cause 
transitions to the molecular states which separate to $^1$S$_0$+$^3$P 
atoms.  The short-range molecular spin-orbit couplings are approximated 
using Table A1 of Ref.~\cite{Hay76}.  The $^1\Sigma_u$-$^3\Pi_u$ and 
$^1\Pi_g$-$^3\Sigma_g$ matrix elements are $\lambda/\sqrt{2}$ and 
$\lambda/2$ respectively, where $\lambda=1.84\times 10^{-4}$ atomic 
units is $2/3$ of the atomic $^3$P$_2$-$^3$P$_0$ splitting.

For the RE studies we assume that the trap depth is 0.5 K, and set the 
probe state to cross the excited quasimolecular state at the point 
$R_e$ where a kinetic energy increase of 1 K for the atom pair has 
been gained after excitation.  Equation~(5) of Ref.~\cite{Julienne94} 
shows that $|S_{gp}|^{2}$ can be factored into two terms: 
$|S_{gp}|^{2}=J_{Q}P_{\rm decay}$, where $J_{Q}$ is the probability of 
being excited from $g$ and reaching $R_e$ on $e$ ($J_{Q}$ accounts for 
the possibility of multiple vibrations).  $J_{Q}$ is calculated from
$|S_{gp}|^{2}$ as explained in Ref.~\cite{Julienne94}, and $P_{\rm decay}$, the
total probability of radiative escape during a vibration in the region
$R<R_e$, is calculated using a classical trajectory:
\begin{equation}
   P_{\rm decay}=1-\exp(-a), \qquad a=2\int_{R_e}^{0}
   dR\frac{\Gamma_{\Sigma_u}(R)}{v(R)} ;
\end{equation}
$\Gamma_{\Sigma_u}(R)$ is given by Eq.~(\ref{Gs}), and $v(R)$ is the
classical speed related to the $^1\Sigma_u$ state potential. 

\begin{figure}[h]
\noindent\centerline{
\psfig{width=80mm,file=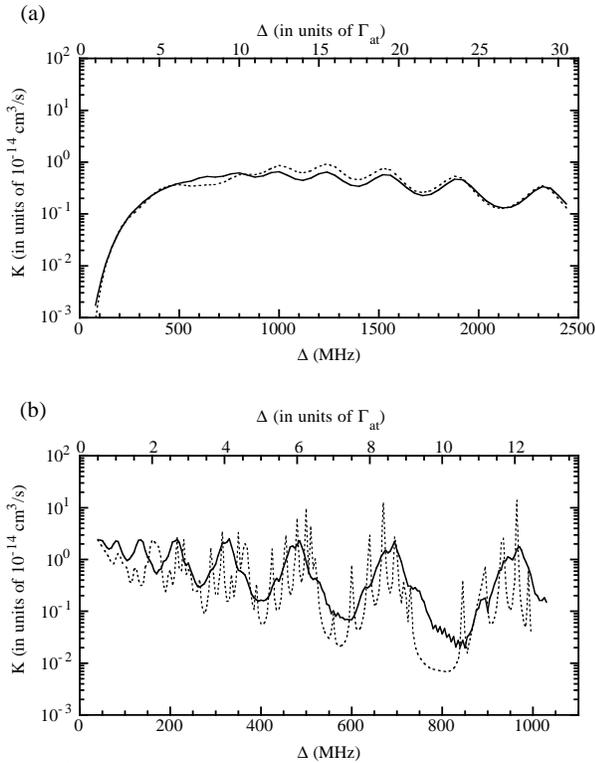}
}
\caption[f2]{The SC loss rate coefficient $K_{\rm SC}$ as a function of laser
detuning $\Delta$ for Mg at laser intensity $I=1$~mW/cm$^2$ and $T=2$ mK
(the Doppler temperature for the chosen Mg cooling transition is $T_{\rm
Doppler}=1.9$ mK). The dotted lines show the sum over branches and partial
waves for $\varepsilon=k_BT$, and the solid lines are the energy averaged
result.  a) $^1\Sigma_u$ state, b) $^1\Pi_g$ state.
\label{SCfig}}
\end{figure}

Fig.~\ref{SCfig} shows calculated rate coefficients $K$ for SC loss 
for both states at laser intensity $I=1$ mW/cm$^2$.  K scales linearly 
with $I$ in the weak field limit.  When $\Delta$ is small, the 
magnitude of K for the $^1\Sigma_u$ state is small and vibrational 
state structure is clearly suppressed even before thermal averaging; 
some structure begins to appear beyond $\Delta \approx 10\Gamma_{\rm at}$. 
However, for the $^1\Pi_g$ state strongly modulated vibrational structure 
persists over the whole range of detuning with comparable peak 
magnitudes.  The $^1\Pi_g$ vibrational structure is broadened but not 
eliminated by the sum over rotational state structure and by thermal 
averaging.  The specific vibrational peak locations versus $\Delta$ 
for both states depend on our particular choice for the excited state 
potentials and may be expected to shift if the correct potentials were 
used.  However, the magnitudes and qualitative nature of the 
oscillations will be independent of this choice.  For very small 
detunings the $^1\Pi_g$ state dominates SC loss, due to much more 
favorable survival after excitation.  For $\Delta\gtrsim 10\Gamma_{\rm 
at}$ $^1\Sigma_u$ and $^1\Pi_g$ contributions become roughly equal.  
The survival probability relative to spontaneous decay at long range 
approaches unity for the $^1\Sigma_u$ state when $\Delta\gtrsim 
10\Gamma_{\rm at}$.  Note that the magnitude of the rate coefficients 
for both $^1\Sigma_u$ and $^1\Pi_g$ states is rather small, on the 
order of $10^{-14}$~cm$^3$/s.  This is because of the relatively small 
spin-orbit matrix elements at the short-range crossings.

%fig 2 here

\begin{figure}[h]
\noindent\centerline{
\psfig{width=80mm,file=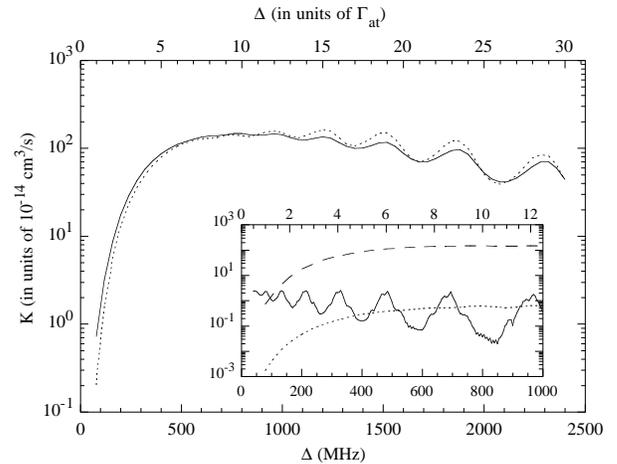}
} 
\caption[f3]{The RE loss rate coefficient $K_{\rm RE}$ for the 
$^1\Sigma_u$ state as a function of the red detuning $\Delta$ for 
Mg at laser intensity $I=1$ mW/cm$^2$ and $T=2$ mK. The trap 
depth is set to 0.5~K. The inset shows the temperature averaged 
results for the loss rate coefficients $K$ for all three loss 
processes (short dash, solid, and long dash lines are from 
Figs. 2a, 2b, and 3 respectively).
\label{REfig}}
\end{figure}

The RE loss for the $^1\Sigma_u$ state, shown in Fig.~\ref{REfig}, 
qualitatively follows that of the corresponding $^1\Sigma_u$ state SC 
loss (see the inset in Fig.~\ref{REfig}), but is about 200 times 
stronger.  Because of this inherently larger magnitude for RE loss, 
trap loss for Mg is predominantly due to the $^1\Pi_g$ state only for 
very small detunings on the order of $\Gamma_{\rm at}$.  Radiative 
escape via $^1\Sigma_u$ excitation is dominant at all larger detunings 
and masks any vibrational structure due to $^1\Pi_g$ SC loss.  
Vibrationally resolved (but not rotationally resolved) photoassociative 
structure should begin to appear due to $^1\Sigma_u$ excitation for 
detunings larger than around 1 GHz, with a rate coefficient on the 
order of $10^{-12}$ cm$^3$/s for $I=1$ mW/cm$^2$.

%fig3 here

Our calculations for Mg shed some light on Sr collisions.  Since the 
spin-orbit coupling strength $\lambda$ is a factor of 9.6 larger for 
Sr than for Mg, the SC loss will be much stronger in Sr than we find 
here for Mg, and thus will compete with RE as a loss mechanism.  If we 
simply scale up our peak magnitude of $^1\Pi_g$ loss at 
$\Delta=4\Gamma_{\rm at}$ linearly in intensity $I$ and by the 
increased $\lambda^2$ factor (differences in mass also need to be 
taken into account), the resulting rate coefficient has the same order 
of magnitude as that which was recently measured~\cite{Dinneen99}.  
This suggests that Sr trap loss at small detuning, unlike that for Mg, 
has a prominent contribution from SC by the $^1\Pi_{g}$ state, 
consistent with the understanding of Ref.~\cite{Dinneen99}.  Since it 
is likely that Sr will have a modulated photoassociation spectrum at 
small $\Delta$ analogous to Fig.~\ref{SCfig}b, measuring trap loss 
versus $\Delta$ would be a very useful experiment to try for the Sr 
system.  Such modulations might also be observed in both Sr and Mg 
by detecting $^3$P$_{0,1,2}$ product atoms, as in similar alkali 
experiments~\cite{Fioretti97,Stwalley98}.

Exploring the alkaline earth systems offers the prospects for some 
very fruitful science.  There is clearly a need for better {\it ab 
initio} or semiempirical potentials and long-range dispersion 
coefficients.  Photoassociation spectroscopy over a wide range of 
detuning values should help characterize ground and excited state potentials 
and ground state scattering lengths~\cite{RMP}.  The recent cooling of 
Sr below 1~$\mu$K using the $^1$S$_0$-$^3$P$_1$ 
transition~\cite{Katori99} raises the possibility of achieving 
Bose-Einstein condensation with an alkaline earth species.  Comparing 
experimental trap loss for alkaline earth isotopes with and without 
hyperfine structure might also shed some light on the role of 
hyperfine structure in alkali trap loss.

This work has been supported by the Academy of Finland, Nordita, NorFA,
and the Office of Naval Research.  We thank Nils Andersen and Jan
Thomsen for discussions.

\end{document}